\documentclass[american,aps,pre,showpacs,twocolumn]{revtex4-2}
\usepackage{graphicx}
\usepackage{amsmath}
\usepackage{color}

\makeatletter
\usepackage{babel}
\usepackage{soul}
\usepackage{ulem}
\makeatother

\begin{document}

	\title{Ballistic File diffusion of Hard-Core particles in One-Dimensional Channels: A Numerical Study.}
	\author{{P. M. Centres$^{1,2,^{*}}$, S. J. Manzi$^{1}$, V. D. Pereyra$^{3}$}, S. Bustingorry$^{4,5,6}$}
	\affiliation{$^1$Departamento de F\'{\i}sica, Instituto de F\'{\i}sica Aplicada, Universidad Nacional de San Luis-CONICET,  Ej\'ercito de Los Andes 950, D5700HHW, San  Luis, Argentina.\\$^2$Universidad Nacional de San Juan, Dpto de Geof\'{\i}sica y Astronom\'{\i}a. Mitre 396 (E), San Juan, J5402CWH, Argentina.\\$^3$ Departamento de F\'{\i}sica, Instituto de Matem\'atica Aplicada San Luis, Universidad Nacional de San Luis-CONICET,  Ej\'ercito de Los Andes 950, D5700HHW, San  Luis, Argentina.\\$^{4}$Instituto de Nanociencia y Nanotecnología (CNEA-CONICET), Nodo Bariloche, Av. Bustillo 9500 (R8402AGP), S. C. de Bariloche, R\'{\i}o Negro, Argentina.\\
    $^{5}$Gerencia de F\'isica,  Centro At\'omico Bariloche, Av. Bustillo 9500 (R8402AGP), S. C. de Bariloche, R\'io Negro, Argentina.\\
    $^{6}$Aragon Nanoscience and Materials Institute (CSIC-University of Zaragoza) and Condensed Matter Physics Department, University of Zaragoza, C/Pedro Cerbuna 12, 50009 Zaragoza, Spain.}

\begin{abstract}

One-dimensional movement of interacting particles is a challenging problem where the correlation between particles induces non-trivial collective effects. In contrast to the single-file diffusion case, the pure ballistic single file movement of particles has received less attention. Here, the ballistic file diffusion of hard disks is studied using an addaptative continuum Monte Carlo numerical scheme. Dynamics is studied as a function of the size of the particles, the system size and the number of particles. The mean square displacement presents three regimes corresponding to independent motion, collective motion and finite size effects. These regimes and the crossover times between them are analyzed and presented in analogy with the ones observed for the single-file diffusion problem.

\textit{Keywords: Diffusion, Hard-Core interaction, Monte Carlo Molecular Dynamics, Mean Square Displacement, Roughness.}
\end{abstract}

\pacs{
	05.60.-k, 
	02.50.Ey, 
	05.70.Np 
}
\date{\today}

\maketitle

\section{Introduction}
\label{introduction}

The concept of single-file diffusion (SFD) comes from physiology \cite{Hodgkin,Taloni,Harris,Arriata,Richards,Fedders,Alexander,Lizana,Beijeren} and  it is a well-known topic in statistical mechanics. Basically, single-file problems study the movement of particles on a narrow channel under the condition that they can not overcome each other. Therefore, SFD is the study of the diffusion of a tagged particle in a single file. This single-file movement is geometrically constrained, the sequence of particles remains the same over time, and the long-time diffusion of an individual particle in a large system is suppressed.

Most single-file problems are charactherized by measuring the mean square displacement (MSD) of each particle. In the general case, the MSD for SFD problems has three regimes as a function of time: For small times an independent diffusive motion is aquired, for longer times particles start interacting with each other resulting in a hindered sub-diffusive regime with the MSD growing as $t^{1/2}$, finally for finite systems when the motion of all particles is correlated the whole system diffuses normally, i.e. the center of mass of the system grows linearly with time~\cite{Beijeren,Lucena}.

Analogously, the ballistic single file (BSF) problem  \cite{Coste,Delfau,Tkachenko} refers to a set of particles with constant velocities moving on a narrow channel. If the repulsive forces among them are hard-core, the interaction reduces to a type single-binary. Under this condition, a collision takes place when two particles are separated at a distance equal to their diameters. Elastic collisions imply that velocities are changed conserving energy and momentum. 

The SFD case, which corresponds to an overdamped dynamics, has been intensively studied. When in addition to the overdamped part the full dynamics is considered, the ballistic contribution to the MSD is expected. Studies considering the full dynamics show that a first ballistic regime is always present \cite{Delfau,Coste}. The subsequent regimes can present both sub-difussion and/or diffusion due to correlated motion of the particles and ballistic or diffusive motion due to finite size effects at very long times.
Therefore both the ballistic and diffusive contributions compete with each other and are present in the full description of the particle movement appearing when using the full dynamics. To cleanly describe the ballistic contributions, detached from overdamping effects, here we present a numerical study of a pure ballistic system in one dimension, allowing to indentify the key signatures of the BSF problem. By using a adaptative Monte Carlo simulation for hard-core particles and using the MSD and the roughness of the system we identify three regimes: independent ballistic motion, normal diffusion, and fully correlated (ballistic or still) dynamics. In addition, we clearly identify the intermediate diffusive regime as due to the cooperative movement of ballistic particles. The long time limit might be ballistic or non-diffusive, depending on whether periodic or close boundary conditions are used, respectively. 

The outline of this paper is as follows: in Sec.~\ref{model} the model and the simulation method are described, and the observables are defined; Sec.~\ref{results} presents the main results of the work and their discussions. Finally, in Sec.~\ref{conclusions} the conclusions are presented.

\section{Model, Simulation Method and Observables}
\label{model}

As it is well known, the numerical simulation of particle motion in the continuum is a very challenging task, particularly due to the fact that it is computationally time-consuming~\cite{Allen,Gillespie,Gim1}.
In Ref.~\cite{Liborio}, a new Markov Chain Monte Carlo method for simulating the dynamics of molecular systems characterized by hard-core interactions has been recently introduced. The state of the system is associated with the set of paths travelled by the atoms and the transition probabilities for an atom to be displaced are proportional to the corresponding velocities. In this way, the number of possible state-to-state transitions is reduced to a discrete set, and a direct link between the Monte Carlo time step and actual physical time are naturally established.
To simulate BSF processes with hard-core interactions in continuous media we use the algorithm proposed in Ref.~\cite{Liborio} with an additional addaptative displacement scheme, as described below. As a consequence, the total execution time is significantly reduced without losing accuracy.


Our model consist of $N$ disk-particles of diameter $\phi$ that are able to diffuse in a channel of length $L$ and fixed wide $\phi$, thus resulting in a one-dimensional constriction. The unit length corresponds to a unitary diameter, $\phi_0=1$, and particles interact via hard-core repulsion only. Two boundary conditions shall be considered: periodic and closed (in a box).
As an initial condition, the particles are located equidistant.
According to Ref.~\cite{Torres} the effect of the initial condition, whether random or flat, on the behavior of observables characterizing SFD is negligible .
In order to implement the Monte Carlo simulation for the diffusion process analyzed here, this initial distribution is preferred to the random initial case because the later is very difficult to obtain, particularly at high densities (note that the jamming coverage for the random case in the continuum is $\approx 0.747598...$ \cite{Evans}).

The velocity $v_i$ of each particle is asigned using a uniform distribution in the interval $-v_0/2 < v_i < v_0/2$. From now on this serves to define reduced time units in terms of $\tau_0 = \phi_0/v_0$.
We shall use $v_0 = 0.1$ in these reduced units.
A key parameter of the simulation is the maximum displacement allowed for a particle in a single movement attemp, which is set to $\delta$.
A displacement $r_{1}\delta$ is calculated for each Monte Carlo step (MCS), where $r_{1}$ is a random number chosen in the $(0, 1]$ interval. Then, the propensities of individual particles are calculated as:
\begin{equation}
	a_{i}=\frac{\mid v_{i} \mid}{r_{1}\delta},
\end{equation}
while the full propensity of the system is
\begin{equation}
	A=\sum_{i=1}^{N} a_{i}.
\end{equation}
The particle, $\mu$, to be displaced is chosen such that it satisfies the following inequality
\begin{equation}
	\sum_{i=1}^{\mu-1}a_{i}<r_{2}A\leq\sum_{i=1}^{\mu}a_{i},
\end{equation}
where $r_{2}$ is a randomly chosen number in the $(0, 1]$ interval.
Then the distance to a pair collision for disk $\mu$, $\delta_{co} = |x_{\mu+\mathrm{sign}(v_\mu)} - x_\mu|$, is calculated along the direction of $v_{\mu}$, keeping all other disks fixed.
On one hand, if $r_{1}\delta <\delta_{co}$ there is no collision, the position is updated to $x_\mu + \mathrm{sign}(v_\mu)r_1 \delta$, and the associated time increase calculated as:
\begin{equation}\label{Taus}
	\tau=\frac{1}{A}.
\end{equation}
 On the other hand, 
if $r_{1}\delta \geq \delta_{co}$ there is a collision with a neighbour particle, the position is updated to $x_\mu + \delta_{co}$
and the associated time is calculated as:
\begin{equation}\label{Taus_ran}
	\tau=\frac{\delta_{co}}{\sum_{i=1}^{N} v_{i}}.
\end{equation}
Following the collision event, the velocities of the colliding particles are updated exactly as in the conventional event driven Molecular Dynamics, i.e., assuming a perfectly elastic collision preserving kinetic energy and momentum~\cite{Haile}. 


Due to the fact that the increments of time $\tau$ are random the time $t$ is actually a mean value obtained over the simulation runs. A discrete set of times is defined in advance, $\left\lbrace t_{0},t_{1},t_{2},...,t_{M}\right\rbrace $, with $t_0 = 0$ and $t_M$ the maximum defined value. While for each run the simulation time is computed according to Eqs.~\eqref{Taus} and \eqref{Taus_ran}, whenever this time is larger than a given $t_j$ for the first time, the time stamp and the corresponding measured observables are save for averaging.

The mean square displacement is defined by
\begin{equation}\label{x2}
	\left\langle x^{2}(t)\right\rangle =\left\langle \frac{1}{N} \sum_{i=1}^{N}  \left[ \Delta x_{i}(t) \right]^{2}\right\rangle,  
\end{equation}
where $ \Delta x_{i}(t) = x_{i}(t)-x_{i}(0)$ gives the displacement with respect to the initial position for the particle $i$, $N$ is the total number of particles, $ \left\langle ... \right\rangle$ represents the average over runs.


In the context of interfaces, the global roughness measures average height deviations from its mean value. Then, considering the displacement of the particles as interface height variables, the single file processes can also be analyzed using the global roughness which thus provides complementary information~\cite{Centres1,Centres2}. It is expected that the time evolution of the correlation length and global roughness have the same scaling properties~\cite{Kolakowska}. The global roughness can be defined by 

\begin{equation}\label{w2}
	\left\langle W^{2}(t)\right\rangle  = \left\langle \frac{1}{N} \sum_{i=1}^{N} \left[\Delta x_{i}(t) - \left\langle \Delta x(t)\right\rangle  \right]^{2}\right\rangle,
\end{equation}
where 
\begin{equation}
	\left\langle \Delta x(t)\right\rangle = \frac{1}{N} \sum_{i=1}^{N} \Delta x_{i}(t)
\end{equation}
gives the average center of mass displacement at time $t$. The global roughness provides thus information about the fluctuations of the particles around the center of mass of the systems.


\begin{figure}
	\centering
	\includegraphics[width=0.48\columnwidth]{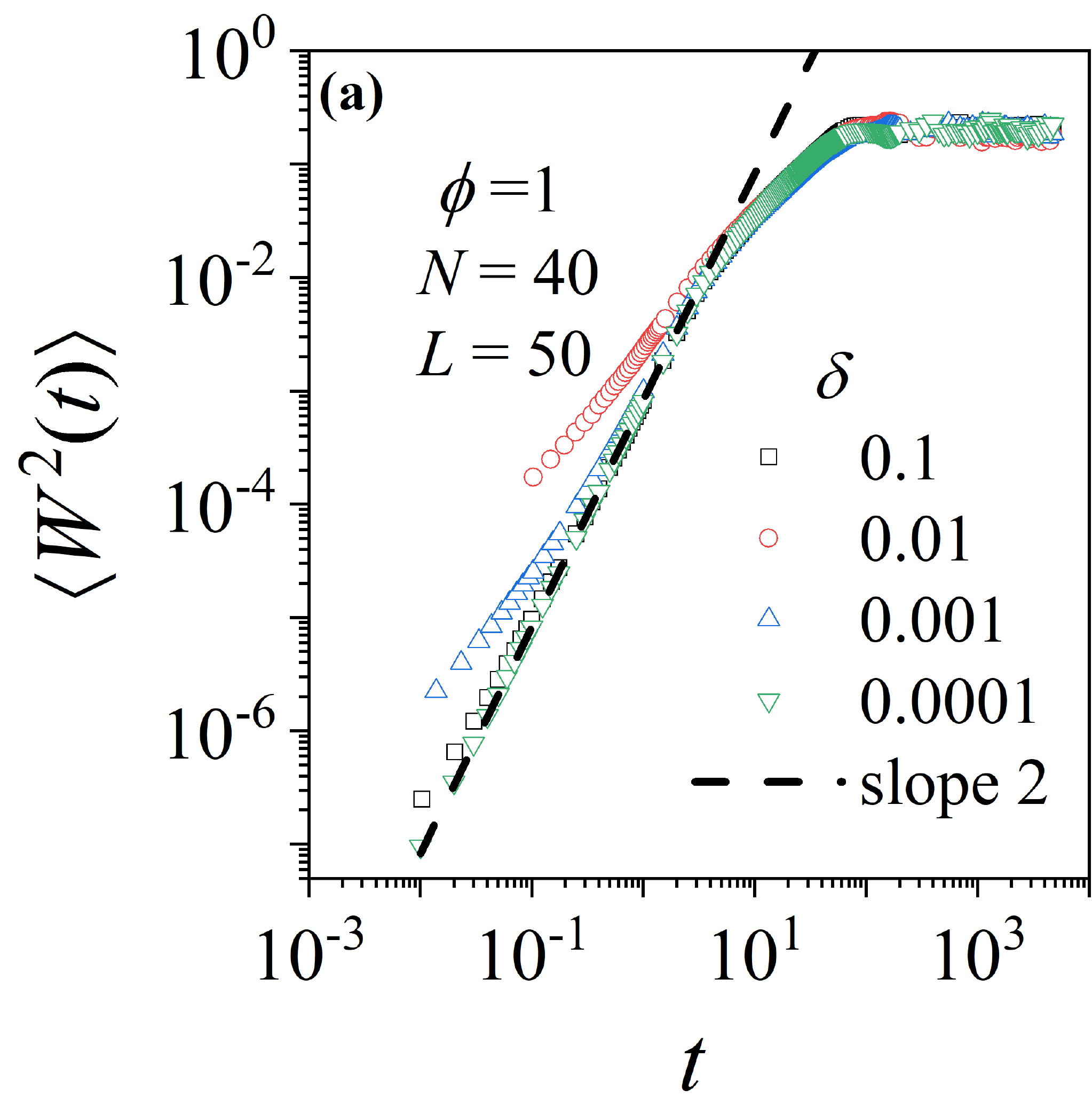}
	\includegraphics[width=0.48\columnwidth]{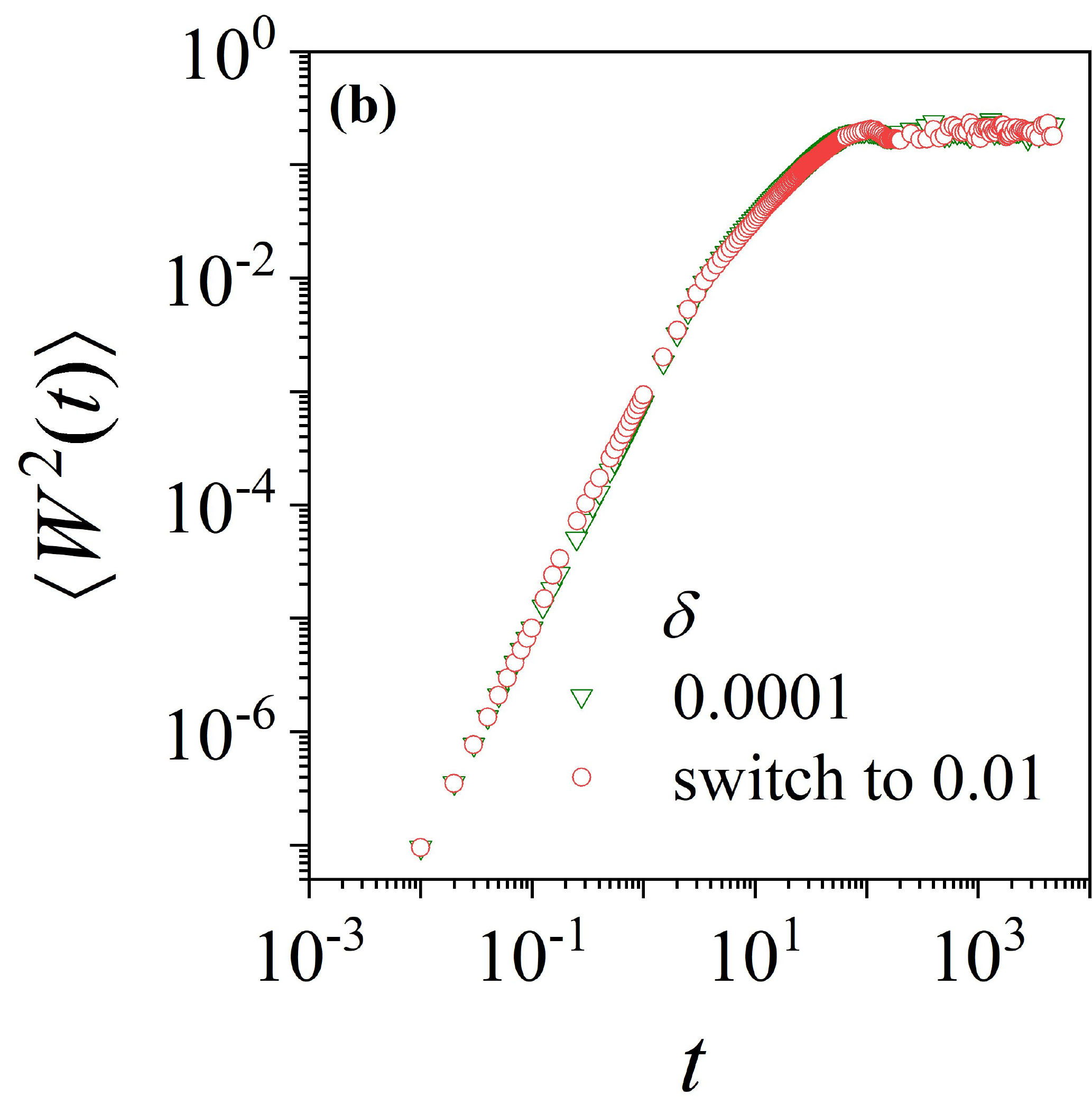}

	\caption{(a) Dependence of the roughness on the maximum allowed displacement $\delta$ in the used algorithm. Small $\delta$ values are needed to accurately describe the initial ballistic regime. (b) Roughness results obtained using an addaptative protocol for the value of $\delta$, resulting in accurate results and shorter computational times.
	}
	\label{Fig01}
\end{figure}

Following the algorithm just described and the details in Ref.~\cite{Liborio}, one can realize that the smaller the maximum allowed displacement $\delta$, the greater the accuracy of the obtained results. In other words, if $\delta \to 0$ the dynamics converges to the deterministic solution. Using a small $\delta$ value and seeking accurate results then implies diverging simulation times. 
%
Figure \ref{Fig01}(a) shows how the choice of $\delta$ affects the measurement of the roughness. Each Curve has been averaged over 100 runs.
As expected, for short times $\langle W^{2} \rangle$ scales with time as $t^{2}$, corresponding to the independent ballistic motion of particles. This regime is well described only when $\delta$ is small enough, while the long time limit is reproducible even for large $\delta$ values. We then adopt and addaptative protocol where all simulation runs initially use $\delta = 0.0001$, which is subsequently changed to a higher value.
Figure~\ref{Fig01}(b) shows such an addaptative simulation run, where $\delta = 0.01$ was used for longer times. All the dynamical regimes are well described by this addaptative numerical approach using feasible simulation times.
To determine the value of $\delta$ and its range of execution, a first running test is performed for each case.
In this way, the execution simulation time is significatively reduced.

%
%

\section{Results and discussion}
\label{results}

\begin{figure}
\centering
\includegraphics[scale=0.35]{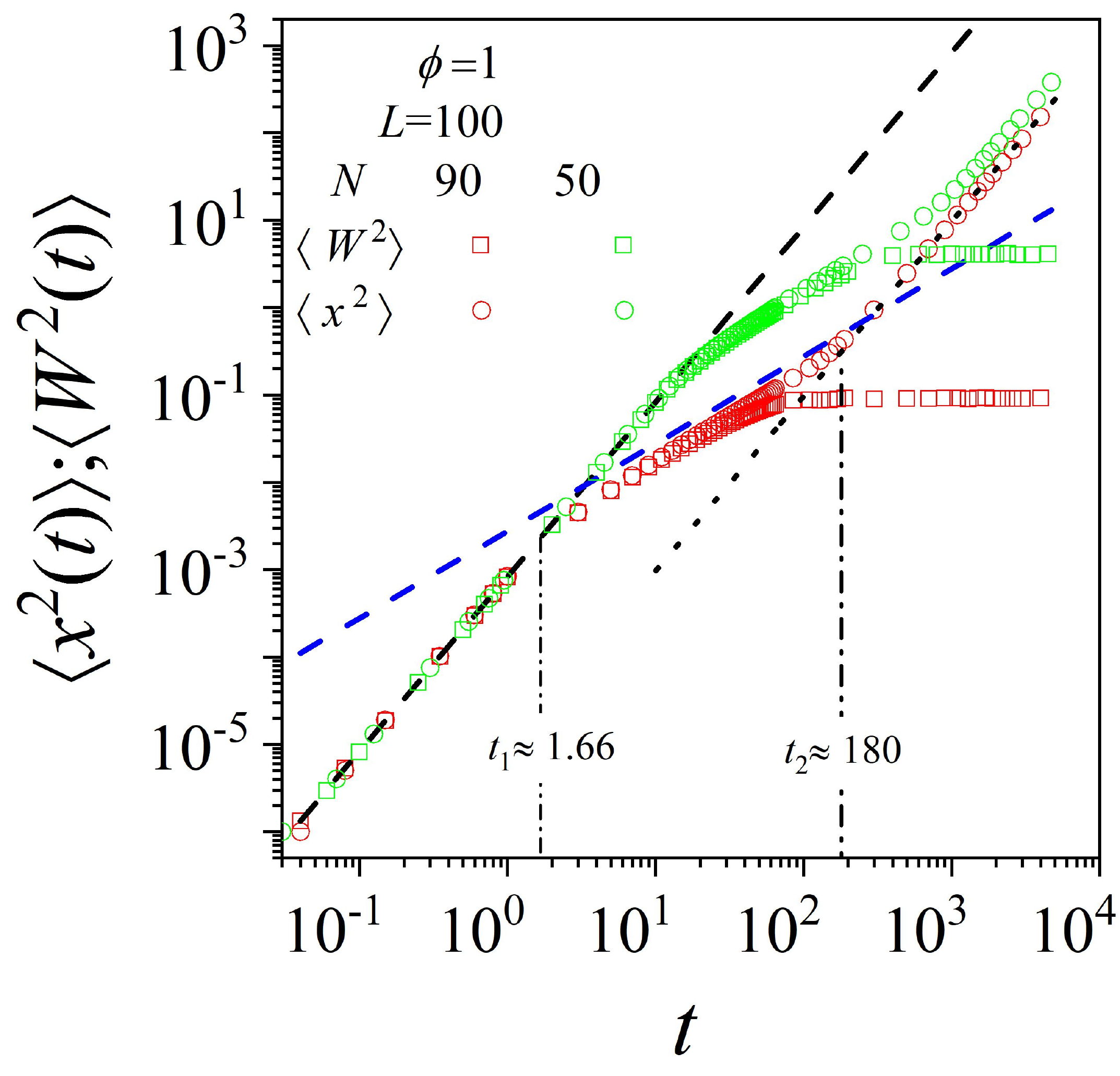}
\caption{Mean-square displacement (circles) and global roughness (squares) for particles with a hard-core interaction obtained through numerical simulations. For $t\rightarrow 0$, The MSD and global roughness are proportional to $t^2$ (black dash line) up to $t_1$. For $t_1<t<t_2$, the MSD and $\langle W^{2} \rangle$ are proportional to $t$ (blue dash line). For $t>t_2$, the mean-square displacement is again proportional to $t^{2}$ (black dot line-fitting) and the roughness reaches a saturation value. Lines correspond to the theoretical approximation Eq.~\eqref{x2}.
}
\label{Fig02}
\end{figure}

\begin{figure}
\centering
\includegraphics[scale=0.3]{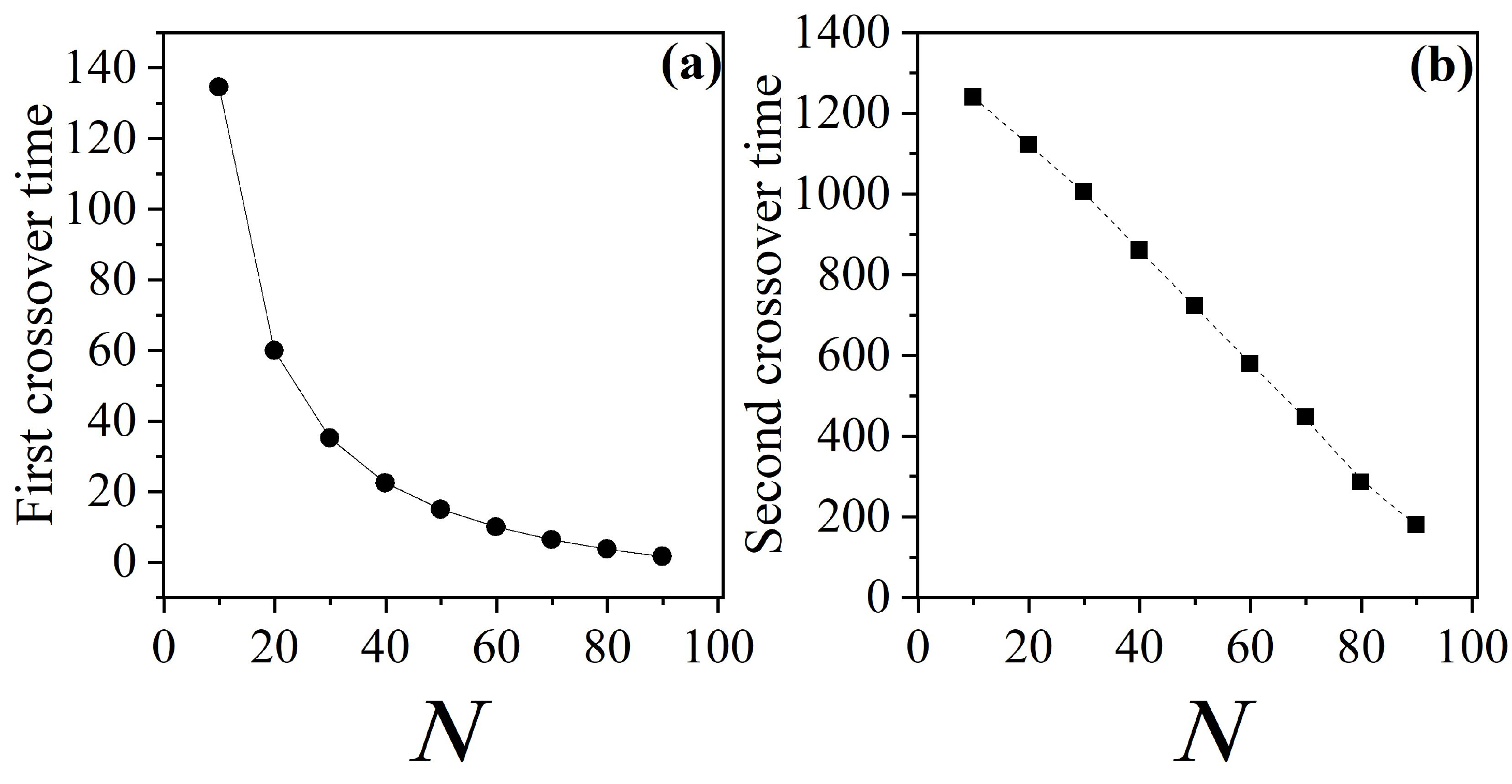}
\caption{Dependence of (a) the first crossover time $t_1$ and (b) the second crossover time $t_2$ on the number of particles in the system, for $\phi=1$ and $L=100$.}
\label{Fig03}
\end{figure}

We shall first describe the three dynamical regimes found using the present BSF model. In Fig.~\ref{Fig02} the MSD (circle symbols) and global roughness (square symbols) are shown as a function of time for a system of size $L=100$, particles of diameter $\phi=1.0$, and two different number of particles $N = 50$ and $N=90$, corresponding to different densities. Periodic boundary conditions are considered. As was mentioned above,  the curves exhibit three different regimes separated by two crossover times, $t_1$ and $t_2$, as indicated in the figure. 


The first regime is characterized by the fact that particles are moving ballistically without colliding with each other. The MSD and global roughness increase with time as $t^{2}$. In the second regime, both MSD and global roughness grow as $t$, indicating that particles are colliding with their first neighboring particles and exchanging their velocities. Finally, in the third regime, MSD again increases as $t^{2}$, meaning that the BSF turns to a regime where particles move as a whole system with a renormalized mass. Instead, the global roughness becomes flat, which means that the correlation length has reached the size of the system, as described in related surface growth problems~\cite{Centres1,Centres2}. This suggest that, for BSF with hard-core interaction, collisions between particles correlate their motion, which implies a growing correlation length as in SFD processes.

The different regimes and the crossovers between them can be rationalized as follows. In the independent ballistic regime the MSD is expected to be $\langle x^2 \rangle = \langle v^2 \rangle t^2$, where $\langle v^2 \rangle$ is the average cuadratic velocity. Using the fact that the initial velocities are uniformly distributed in the range $(-v_0/2, v_0/2)$, it results $\langle v^2 \rangle = v_0^2/12$.
In the intermediate diffusive regime $\langle x^2 \rangle = D t$, where the diffusion constant $D$ can be thought of as given by the typical velocity of the particles $\langle |v| \rangle$ times the mean free path between nearest particles $(L - N \phi)/N$, resulting in $D = \langle |v| \rangle (L/N - \phi)$, with $\langle |v| \rangle = v_0/4$.
The long time regime corresponds to the collective ballistic diffusion of the particles and then $\langle x^2 \rangle = \langle v^2 \rangle t^2/N$. Therefore, one can write for the time evolution of the MSD for the BSF that
\begin{equation}\label{x2}
	\langle x^2 \rangle  = \left\{ \begin{array}{ll}
		\vspace{0.25cm}
		\frac{v_0^2}{12} t^2, & \mbox{for $t < t_1$}, \\
		\vspace{0.25cm}
		\frac{v_0}{4} \left( \frac{L}{N} - \phi \right) t, & \mbox{for $t_1 < t < t_2$}, \\
		\frac{v_0^2}{12N} t^2, & \mbox{for $t_2 < t$}.
	\end{array}
	\right.
\end{equation}
From this, the crossover times can be obtained by equating the MSD for different regimes, though their values are slightly overestimated due to the rough approximation to obtain the intermediate regime, as can be observed in Fig.~\ref{Fig02}. In general, one expects that $t_1 \sim (L/N-\phi)/v_0$ and $t_2 \sim N(L/N-\phi)/v_0$, implying $t_2/t_1 = N$.

%
%

The results presented in Fig.~\ref{Fig02} correspond to $\phi=1$, $L = 100$ and different number of particles $N=50$ and $N=90$, and agree with the scaling proposed for the MSD. From the numerical simulations, estimated values for the crossover times can be obtained, as shown in Fig.~\ref{Fig02}. Figure~\ref{Fig03} presentes $t_1$ (a) and $t_2$ (b) as a function of the number of particles $N$, in agreement with $t_1 \sim 1/N$ and $t_2 \propto N(L/N-\phi)= (L-N\phi) \sim -N$.


\begin{figure}
	\centering
	\includegraphics[scale=0.35]{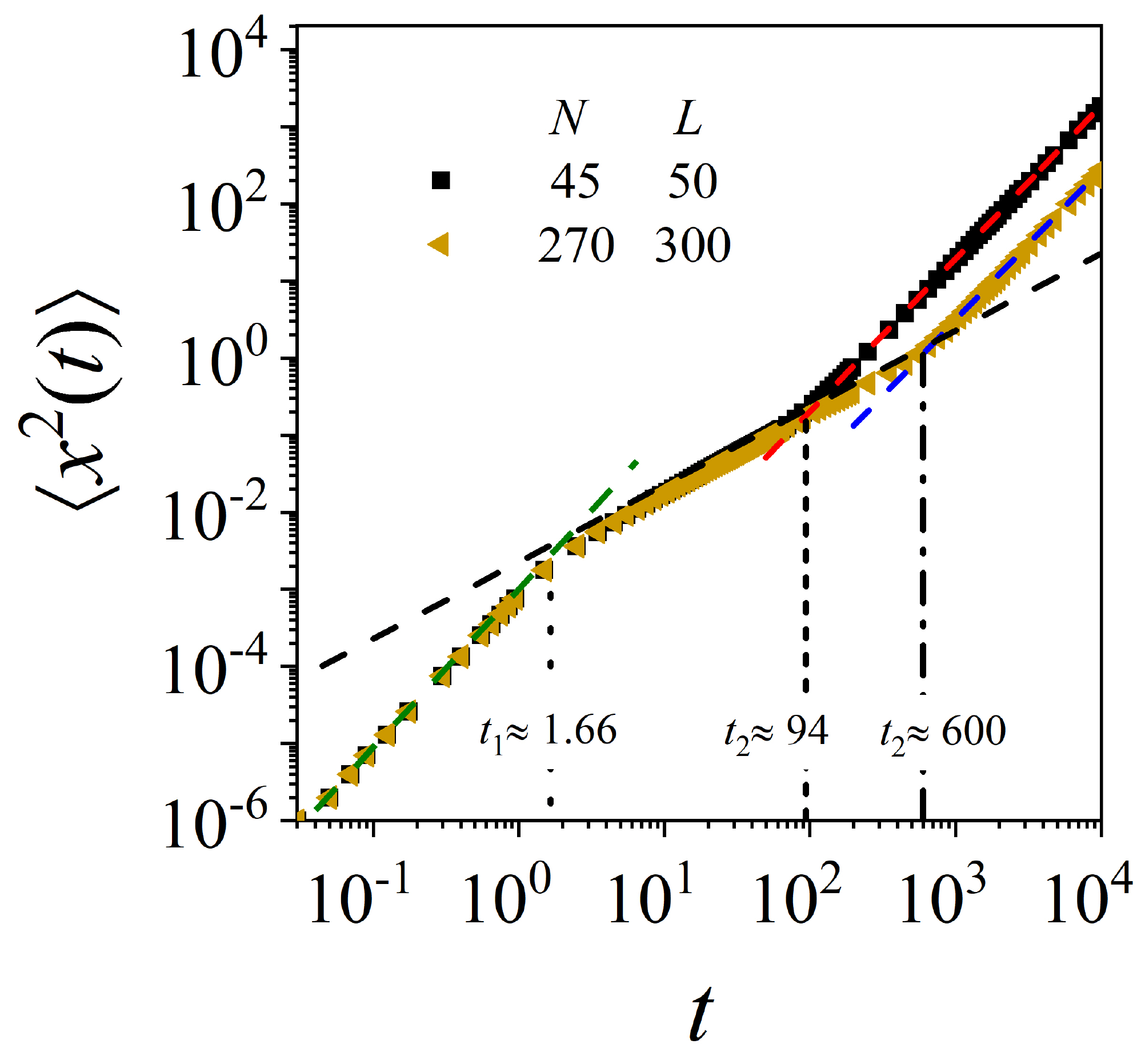}
	\caption{MSD and roughness for $\phi=1$ and for a constant value of $N/L=0.9$. As suggested in Eq.~\eqref{x2}, $t_1$ is not changing in this case while $t_2$ is proportional to $N$.
	}
	\label{Fig04}
\end{figure}

The MSD for constant $N/L=0.9$ (using $L=50$, $N=45$ and $L=300$, $N=270$) is shown in Fig.~\ref{Fig04}. It can be observed that the first crossover time is the same in both cases, as expected from Eq.~\eqref{x2}. On the contrary, the second crossover depends on $N$. This behaviour is in agreement with that shown in Refs.~\cite{Delfau,Tkachenko}. 
 
\begin{figure}
	\centering
	\includegraphics[scale=0.35]{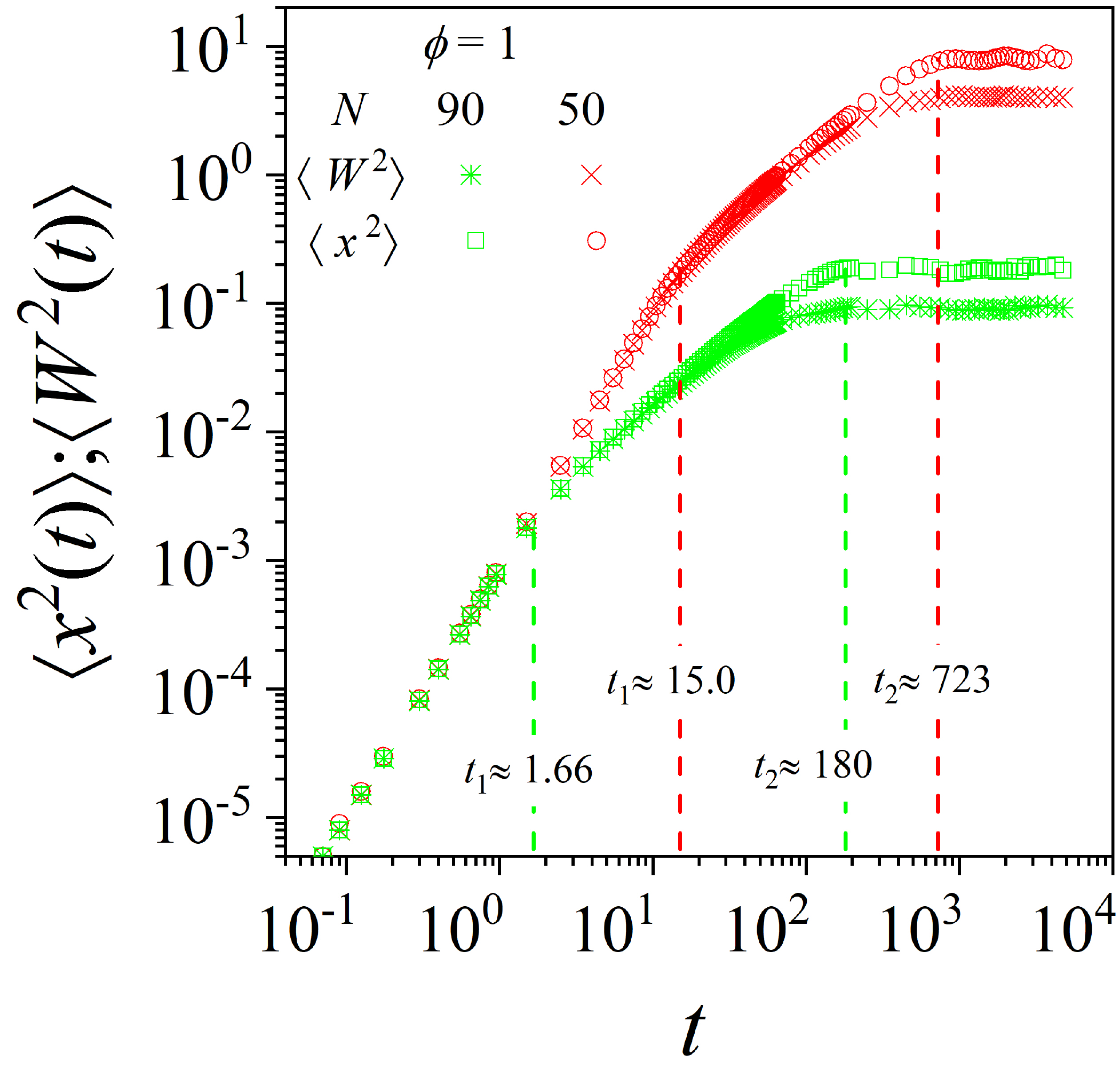}
	\caption{Symbols show the mean-square displacement (circles) and the global roughness (squares) for simulations of particles with a hard-core interaction in a closed box.}
	\label{Fig05}
\end{figure}

Next, we present results of the time evolution of MSD and global roughness in a system with closed boundary conditions, i.e. a closed box, as opposed to periodic boundary conditions. Figure~\ref{Fig05} shows the MSD and global roughness as a function of time for a system of particles of diameter $\phi=1.0$, with size $L=100$ and $N = 50$ and $N = 90$. The results are to be compared with those of Fig.~\ref{Fig02}.  It can be observed that both set of curves show three regimes with roughly the same crossover times as for periodic boundary conditions, but in this case the MSD saturates as the center of mass diffusion is also restricted to the box~\cite{Lizana,Beijeren}. The saturation value of the global roughness is the same as for periodic boundary conditions, indicating that fluctuations depends only on the linear size of the simulated sytem and not on the boundary conditions.




\begin{figure}
	\centering
	\includegraphics[scale=0.35]{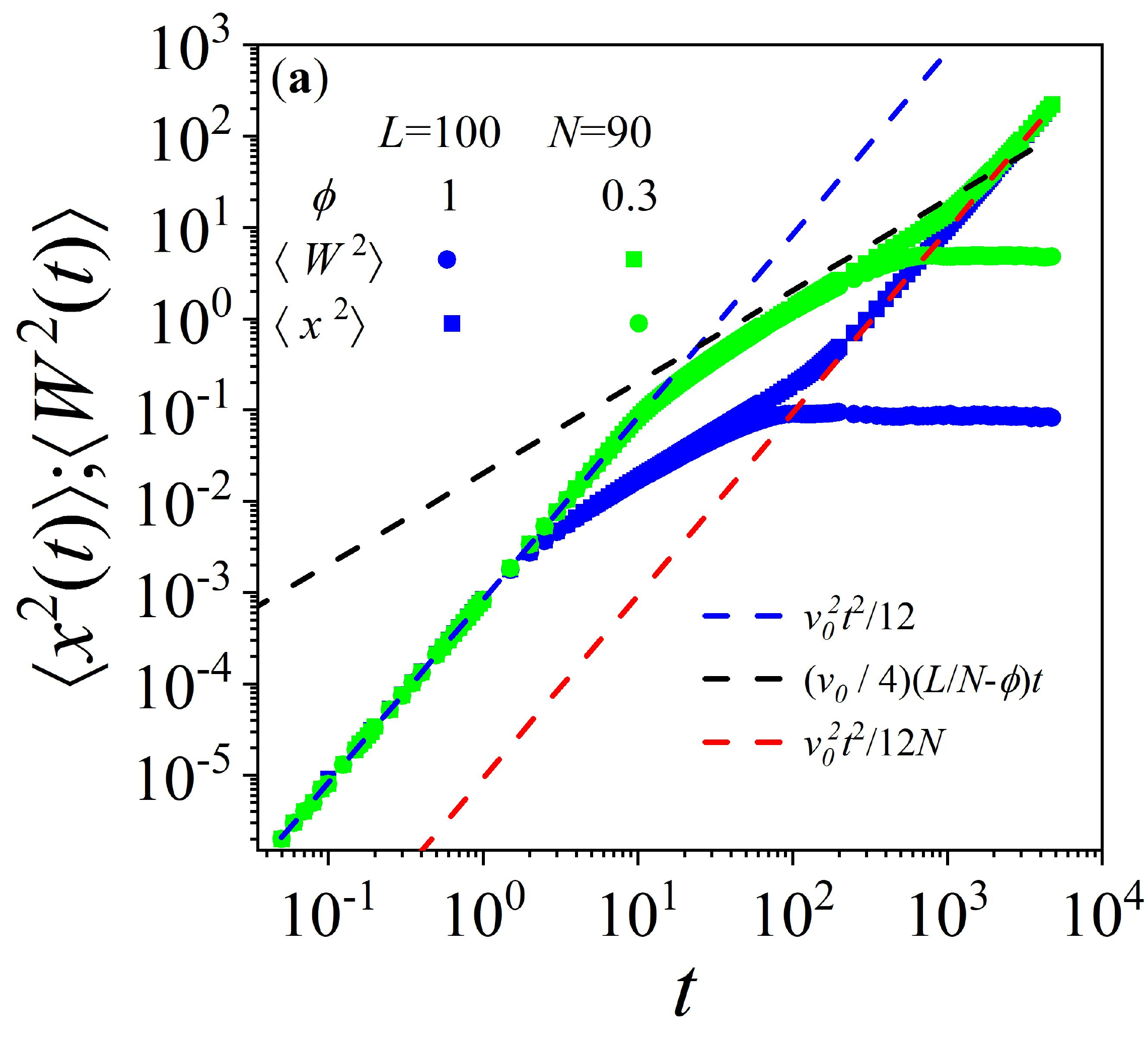}\\
	\includegraphics[scale=0.35]{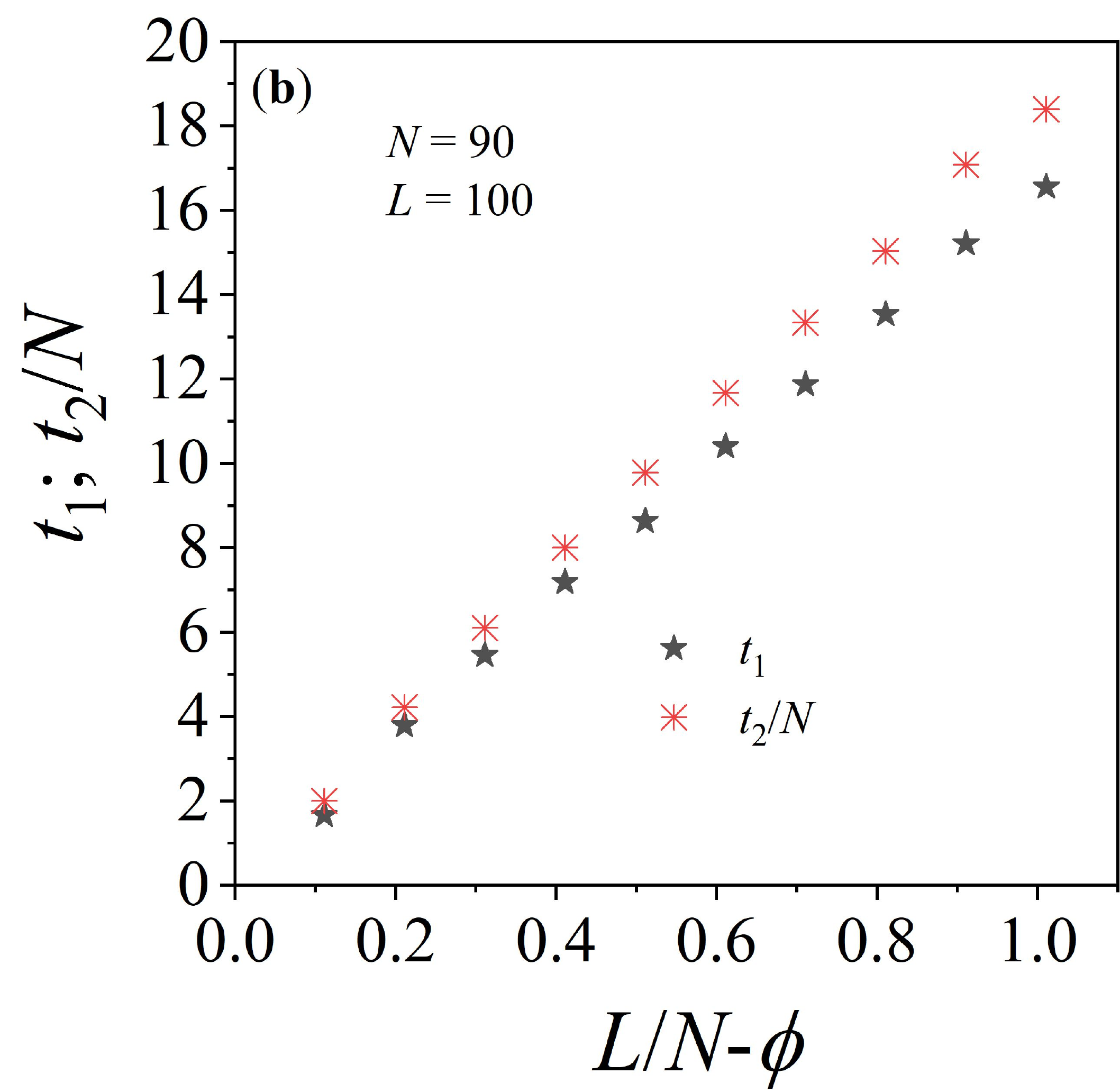}
	\caption{(a) Mean square displacement and $\langle W^{2} \rangle$ for $L=100$ and for two values of $N=50$ and $N=90$ and  two sizes of particles $\phi=1$ and $\phi=0.3$. (b) Shows the dependence of the crossover times as a function of $L/N-\phi$, as suggested in Eq.~\eqref{x2}.
	\label{Fig06}}
\end{figure}
Figure~\ref{Fig06} (a) presents results obtained using different diameters values $\phi=1$ and $\phi=0.3$ for $N/L=0.9$. Periodic and closed boundary conditions are used.
For particles with $\phi=0.3$ the crossover times are larger than those corresponding to $\phi=1$. That is because when the particle diameter decreases while keeping $N/L$ constant, the time between particle collisions increases. Figure~\ref{Fig06}(b) presents the crossover times as a function of $L/N-\phi$, showing a linear increase as suggested in Eq.~\eqref{x2}.




\begin{figure}
	\centering
	\includegraphics[scale=0.35]{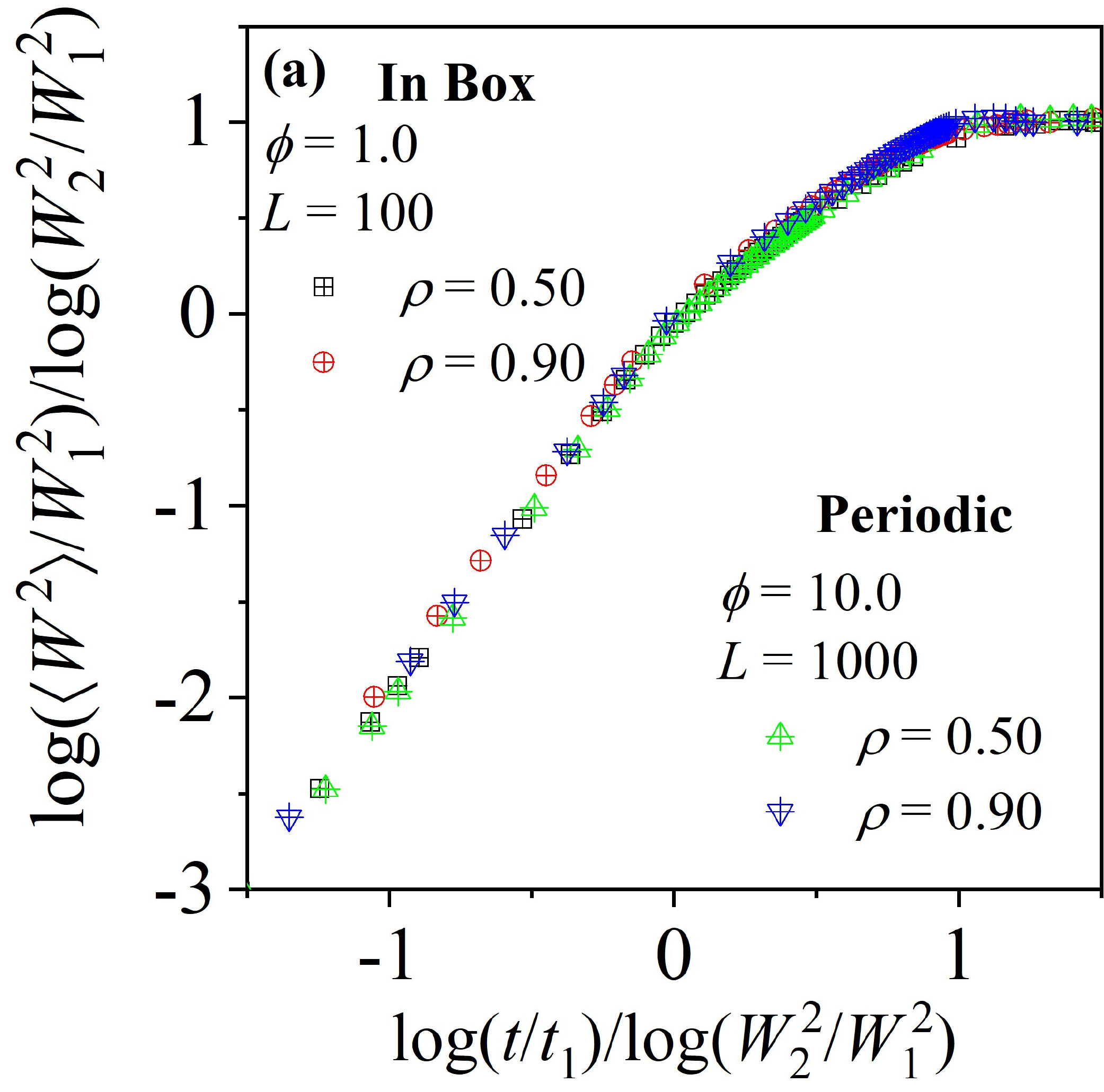}
	\includegraphics[scale=0.35]{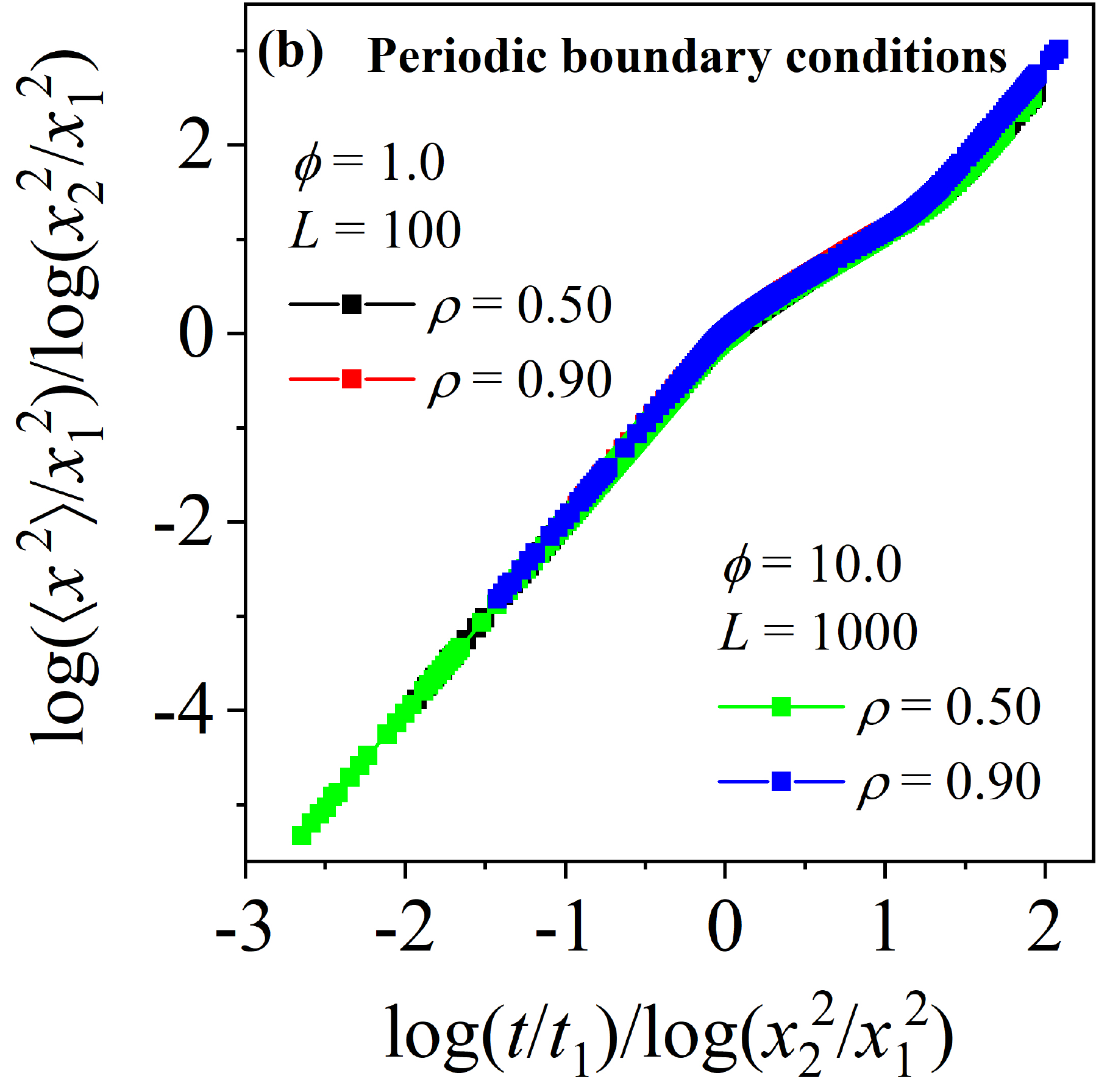}
	\caption{ (a) The phenomenological collapse proposed in Ref.\cite{Chou} for global roughness corresponding to a system of particles of diameter $\phi=1$ and $\phi=10$ with periodic boundary conditions and in a box. (b) The collapsing curves for MSD corresponding to the same system parameters but only for periodic boundary conditions.}
	\label{Fig07}
\end{figure}

Following the phenomenological scaling proposed by Chou and Pleimling \cite{Chou}, once the two crossover points for each curve are estimated, the scaling can be described through:

\begin{equation}
	\frac{\log\left( \frac{\langle W^{2} \rangle}{W^{2}_{1}}\right) }{\log\left( \frac{W^{2}_{2}}{W^{2}_{1}}\right) }=F\left[ \frac{\log\left( \frac{t}{t_{1}}\right) }{\log\left( \frac{W^{2}_{2}}{W^{2}_{1}}\right) }\right],
\end{equation}
where $F(x)$ is a given scaling function containing the three regimes and $W^2_{1,2} = \langle W^2(t_{1,2}) \rangle$. A similar scaling applies to the MSD.
In Figure \ref{Fig07} (a) the collapsing curves corresponding to $\langle W^2 \rangle$ for a systems with two different bondary condition, periodic and in a box are shown. The systems parameters are given in the figure.
The same geometric scaling applies to the BSF process for both the roughness and MSD and exhibits a remarkable collapse, as shown in Fig. \ref{Fig07}.
Note that the collapse does not depend on the boundary condition. In Fig.~\ref{Fig07} (b) the collapsing curves corresponding to MSD for the same system and periodic boundary condition are shown. In analogy with SFD problem, it is oberved that $\langle W^2 \rangle$ in the BSF problem presents the three characteristic time regimes described above.
 
The geometric collapse proposed by Chow and Pleimling~\cite{Chou} for the $\langle W^2 \rangle$ curves was used for the SFD problem in Ref.~\cite{Centres1}, which gives exponents of growth and saturation that coincide with those corresponding to the Edward-Wilkinson universality class. Moreover, when the external field is included, they change to the Kardar-Parisi-Zhang universality class \cite{Centres2}. That is, these scalings have physical meaning and could suggest that the BSF processes are connected with the dynamic scaling of Family and Vicsek, although it is not clear the corresponding universality class.
Indeed, associating this collapse to a dynamic Family-Vicsek scaling is a very difficult task, since the roughness exponent could not be readily determined because the saturation roughness $W^2_2$ depends on the density and the size of the system simultaneously.

%
%
%
%

\section{Conclusions}
\label{conclusions}

In this work, an addaptative continuum Monte Carlo algorithm has been used to study the BSF problem for hard-core particles. This allows us a considerable reduction of the execution times of the numerical simulations while at the same time keeping the accuracy and precision in the calculation of the observables. We thus study the clean BSF problem detached from overdamping effects. In analogy with the SFD problem, we observed three dynamical regimes corresponding to independent ballistic motion, correlated normal diffusion of ballistic particles, and a finite size regime ruled by the center of mass, ballistic or still, depending on the boundary condition. We rationalized these regimes and presented how they depend on the parameters of the system (particle size, system size, and number of particles). Unlike the SFD problem, there is not a direct association of the BSF process to dynamical universality classes, a fact that deserves further attention.


\section{Acknowledgements}
In memory of Professor Giorgio Zgrablich. This work is partially supported by the CONICET-Argentina.

\end{document}